\documentclass[11pt]{article}
\usepackage[latin1]{inputenc}
\usepackage[english]{babel}
\usepackage[namelimits]{amsmath}
\usepackage{amssymb}
\usepackage{amsmath}
\usepackage{amsthm}

\begin{document}
\title{Statistical mechanics of gravitons in a box and the black hole entropy}
\author{
Stefano Viaggiu,\\
Dipartimento di Matematica,
Universit\`a di Roma ``Tor Vergata'',\\
Via della Ricerca Scientifica, 1, I-00133 Roma, Italy.\\
E-mail: {\tt viaggiu@axp.mat.uniroma2.it}}
\date{\today}\maketitle
\begin{abstract}
\noindent This paper is devoted to the study of the statistical mechanics of trapped gravitons obtained by 'trapping' a 
spherical gravitational wave
in a box. As a consequence, a discrete spectrum dependent on the Legendre index $\ell$ similar to the harmonic oscillator one is obtained and a statistical study is performed. The mean energy 
$<E>$ results as a sum of two discrete Planck distributions with different dependent frequencies. As an important application, we derive the 
semiclassical Bekenstein-Hawking entropy formula for a static Schwarzschild black hole by only requiring that the black hole internal energy
$U$ is provided by its ADM rest energy, without invoking particular quantum gravity theories. 
This seriously suggests that the interior of a black hole can be composed of trapped gravitons at a 
thermodynamical temperature proportional by a factor $\simeq 2$ to the horizon temperature $T_h$.

\end{abstract}

\section{Introduction}
The relativistic study of the radial oscillations and of the related instability of a star have begun with the works of Chandrasekhar \cite{1,2}. 
The radial oscillations of a star are not related to the emission of gravitational waves, that are intrinsically quadrupolars. The emission of gravitational waves is related to the so called non-radial oscillations. 
The theory of the non-radial oscillations of a spherical star started with the work in \cite{3} and improved in \cite{4,5}. In 
\cite{6} Chandrasekhar presented a complete relativistic theory of the non-radial oscillations of a black hole as a problem of resonant scattering. There \cite{6}, an incoming gravitational wave perturbates the black hole. The incident wave is in part absorbed and in part reflected by the 
event horizon: the quasi-normal frequencies are obtained by imposing no outgoing waves at spatial infinity. 
The necessary mathematical formalism to 
expand a generic perturbation in a tensorial basis of $10$ spherical harmonics can be found in
\cite{0,01}. This technology has been applied by Chandrasekhar and Ferrari \cite{7,8} and by Ferrari and collaborators (see for example \cite{10,11})
to study the non-radial oscillations of a star.
In particular, the 
equations governing the perturbations of the gravitational field ({\it axial}) outside the star reduce \cite{0,6} to a one-dimensional Schrodinger-like wave equation named Regge-Wheeler equation, while the ones governing the perturbations of the matter fluid outside the star 
({\it polar}) reduce to the Zerilli \cite{01} equation. The presence of a Schrodinger-like equation for stationary states in the theory of non-radial oscillations is an intriguing fact that can be the starting point of some reasonings concerning a quantum description of a gravitational waves in terms of gravitons.

As pointed in \cite{14}, thanks to the intuitions of the founding fathers of quantum mechanics,
the theory of free electromagnetic waves have played a fundamental role 
for the quantization of the free electromagnetic field and for the Planck derivation of the black body radiation. 

For the complexity of the 
general relativity equations, the same has not
happened for the gravitational waves. Moreover,
an electromagnetic radiation can be confined in thermal equilibrium within a cavity (black body radiation), but the same for gravitons is a complicated task. Confining a gravitational radiation in a box is thus a complicated but necessary task in order to explore fundamental 
quantum properties of the gravitational waves. In \cite{15} S. W. Hawking, in order to study the stability of a black hole thermodynamically, 
confined a black hole in a fancy box. In \cite{16} it is noticed that a graviton cannot be trapped in a box. Nevertheless, an embedding
within an anti de Sitter spacetime can make the job, although in a idealized way. In \cite{17,18}, L. Smolin concludes that cannot exist a 
realistic substance capable to absorb a gravitational radiation, while D. Garfinkle and R.M. Wald \cite{19} presented a counter-example of
\cite{17,18} criticized in \cite{20}. Finally, in \cite{21} T. Padmanabhan and T.P. Singh presented
an interesting example to confine a linearized gravitational radiation
in thermal equilibrium within a cavity. The first goal of this paper is to obtain a physically motivated formula for the spectrum
of gravitons trapped in a box, after that  
we study the statistical mechanics of gravitons in a box ('hard wall')  
and some possible applications to the confining gravitons procedure. In particular, we are able to explain the 
semiclassical Bekenstein-Hawking 
black hole formula $S_{BH}=K_B A_h/L_P^2$ (with $K_B$ the Boltzmann constant, $L_P$ the Planck length and $A_h$ the proper area of
the event horizon) for a Schwarzschild black hole
starting from the discrete spectrum of gravitons confined within the black hole.

In section 2 we rewrite the Regge-Wheeler and Zerilli equations in the special case of a perturbed Minkowskian spacetime. 
In section 3 we obtain a discrete spectrum for gravitons in a spherical box, 
while section 4 is devoted to the statistical description of 'trapped' gravitons. In section 5 we analyze the thermodynamics of trapped
gravitons. In section 6 we apply the machinery of sections above to obtain the semi-classical black hole entropy formula 
in a simple and clear way, while in section 7 the case with summation over the legendre index $m$ is done. 
Finally, section 8 is devoted to some conclusions and final remarks.  

\section{The equations in the vacuum}

The first goal of this paper is to obtain a physically motivated expression for the spectrum of gravitons trapped in a 
spherical box. It is customary (see \cite{E1} and references therein) in perturbative quantum field theory,
to depict graviton field $h_{\mu\nu}$ as a perturbation about the flat metric $diag(1,-1,-1,-1)$.
To this purpose, it is sufficient to consider a gravitational wave traveling in a Minkowski
(perturbed) spacetime. Hence,
we consider perturbations of the Minkowski spacetime (we initially use geometrized unity with $G=c=1$)
\begin{equation}
ds^2=dt^2-r^2\sin^2\theta\;d{\phi}^2-dr^2-r^2 d{\theta}^2.
\label{1}
\end{equation}
As usual, we indicate with $g_{ik}^{(0)}$ the unperturbed Minkowskian metric given by (\ref{1}) and with $h_{ik}$ a small perturbation with
$|h_{ik}|<<|g_{ik}^{(0)}|$. The perturbed metric is
\begin{equation}
g_{ik}=g_{ik}^{(0)}+h_{ik}.
\label{2}
\end{equation} 
Since of the spherical symmetry of the 'bare' unperturbed metric $g_{ik}^{(0)}$, the perturbating term $h_{ik}$ can be expanded in a basis of spherical tensorial harmonics \cite{0,01} depending on the real spherical harmonics 
$Y_{\ell m}(\theta,\phi), \ell\in\mathbb{N}
, m\in\mathbb{Z}
, m\in [-\ell, +\ell]$. In this frame, the polar perturbations 
$h_{ik}^{p}$ are the ones dependent on harmonics that under parity operator look like ${(-1)}^{\ell}$, while the axial perturbations
$h_{ik}^{a}$ look like ${(-1)}^{\ell+1}$.\\ 
In the diagonal gauge \cite{10} the perturbed metric becomes, for the axial case:
\begin{equation}
h_{ik}^a=
\begin{pmatrix}
(t) & (\phi) & (r) & (\theta)\\
0 & h_0(t,r)\sin\theta\;Y_{\ell m,\theta} & 0 & -h_0(t,r)\frac{1}{\sin\theta}Y_{\ell m,\phi}\\
h_0(t,r)\sin\theta\;Y_{\ell m,\theta} & 0 & h_1(t,r)\sin\theta\;Y_{\ell m,\theta} & 
0\\
0 & h_1(t,r)\sin\theta\;Y_{\ell m,\theta} & 0 & -h_1(t,r)\frac{1}{\sin\theta}\;Y_{\ell m,\phi}\\
-h_0(t,r)\frac{1}{\sin\theta}Y_{\ell m,\phi} & 0 & -h_1(t,r)\frac{1}{\sin\theta}\;Y_{\ell m,\phi} &
0
\end{pmatrix},
\label{7}
\end{equation}
and for the polar case:
\begin{equation}
h_{ik}^p=
\begin{pmatrix}
(t) & (\phi) & (r) & (\theta)\\
2N(t,r) Y_{\ell m} & 0 & 0 & 0\\
0 & -2r^2\sin^2\theta\;H_{11}(t,r,\theta,\phi) & 0 & -r^2 V(t,r) X_{\ell m}\\
0 & 0 & -2L(t,r) Y_{\ell m} & 0\\
0 & -r^2 V(t,r) X_{\ell m} & 0 & -2 r^2 H_{33}(t,r,\theta,\phi)
\end{pmatrix}, 
\label{8}
\end{equation}
where
\begin{eqnarray}
X_{\ell m}(\theta,\phi) &=& 2Y_{\ell m,\theta,\phi}-2\cot\theta\;Y_{\ell m,\phi}\label{8b}\\
W_{\ell m}(\theta,\phi) &=& Y_{\ell m,\theta,\theta}-\cot\theta\;Y_{\ell m,\theta}-
\frac{1}{\sin^2\theta}\;Y_{\ell m,\phi,\phi}\nonumber\\
H_{11}(t,r,\theta,\phi) &=&TY_{\ell m}+\frac{V}{\sin^2\theta}\;Y_{\ell m,\phi,\phi}+
V\cot\theta\;Y_{\ell m,\theta}\nonumber\\
H_{33}(t,r,\theta,\phi) &=&TY_{\ell m}+VY_{\ell m,\theta,\theta}.
\end{eqnarray} 
As usual, the static nature of the unperturbed metric (\ref{1}) allows to assume that the perturbation functions all have the time dependence
$e^{-\imath\omega t}$ \cite{6} (with $\omega$ a constant): this is equivalent to a Fourier analysis of the metric coefficient with 
frequency $\omega$. Since we are considering vacuum perturbations, we have
$\delta T_{ik}=0$.

 \subsection{Axial perturbations with $\ell\geq 2$}

As well known \cite{0,6}, after defining the Regge-Wheeler function 
$Z_{\ell m}^{(a)}$ by $h_{1\ell m}=rZ_{\ell m}^{(a)}$, the axial perturbations are driven by
the Regge-Wheeler\footnote{We named the (\ref{23}) Regge-Wheeler equation for simplicity, although it refers to
	the Schwarzschild case.} equation \cite{14}:
\begin{equation}
Z_{\ell m,r,r}^{(a)}+{\omega}^2 Z_{\ell m}^{(a)}=\frac{\ell(\ell+1)}{r^2}Z_{\ell m}^{(a)}.
\label{23}
\end{equation}
Equation (\ref{23}) reminds the Schrodinger equation for stationary states with vanishing potential $V(r)$. 

\subsection{Polar perturbations with $\ell\geq 2$}
By following the same manipulations of \cite{14}
and after defining $Z_{\ell m}^{(p)}=rL_{\ell m}$, the 
solutions of the linearized Einstein's equations become:
\begin{eqnarray}
& & qV_{\ell m}(r,\omega)=-L_{\ell m}(r,\omega)-\frac{1}{r}\int_0^r L_{\ell m}(r^{\prime},\omega)dr^{\prime},\label{f1}\\
& & N_{\ell m}(r,\omega)=L_{\ell m}(r,\omega)+2r\int_0^r\frac{L_{\ell m}(r^{\prime},\omega)}{{r^{\prime}}^2}dr^{\prime},
\label{f2}\\
& & Z_{\ell m,r,r}^{(p)}-\frac{\ell(\ell+1)}{r^2}Z_{\ell m}^{(p)}+{\omega}^2 Z_{\ell m}^{(p)}=0.\label{f3}
\end{eqnarray}
To obtain the equation (\ref{f1}) and (\ref{f2}) we have imposed the regularity of $L_{\ell m}$ for $r\rightarrow 0$.

\section{Gravitons in a box}

Starting from this section, we restore the constants $c$ and $G$.
In the literature of statistical black hole entropy, we can often see gravitons inside the horizon depicted as an
ensemble of $n$ non-interacting massless particles
with (continuum) angular frequency \cite{E2} $\omega$ or by considering discrete ad hoc expressions for $\omega$ like 
$\omega\sim n$ or inspired \cite{E3,CC1,CC2,CC3} 
by quasi-normal black hole frequencies. However, although the main proposals present in the literature are physically possible, a derivation from the onset
\footnote{A convincing quantum gravity theory is still not at our disposal.} 
of the (discrete) expected frequency of trapped gravitons is still lacking.
Hence, our primary goal in this section is to 
obtain a physically reasonable ('phenomenological') formula for the frequency of trapped gravitons motivated by the techniques of perturbative quantum field 
theory (see \cite{E1} and references therein), i.e. perturbation around a Minkowski spacetime.

To start with,
as well known, \cite{0,01,6,14} equations (\ref{23}) and (\ref{f3}) governing the axial and polar perturbations respectively look like the radial part of the Schrodinger equation for free particles and for stationary states. The only difference is that to the metric ('wave') function 
$L_{\ell m}$ are associated, as the quantum mechanical case, the Legendre polynomials $Y_{\ell m}(\theta,\phi)$, while for the axial case to
$Z_{\ell m}^{(a)}$ are associated the Legendre polynomials given by $Y_{\ell m,\theta}$ and $Y_{\ell m,\phi}$ (see \cite{7}). This means that the analogy with the Schrodinger equation is more stringent with the polar case. Hence, in the following we only consider equation (\ref{f3}),
although the following considerations concerning the (\ref{f3}) obviously still hold for the (\ref{23}).

As well known, a particle with rest mass $m$, 
potential $V(r)$ and energy $E$ satisfies the spherical Schrodinger equation with a wave 
function ${\psi}_E(r,\theta,\phi)$ that can be split as
${\psi}_E(r,\theta,\phi)=R(r)Y_{\ell m}(\theta,\phi)$. The angular part $Y_{\ell m}$ satisfies the eigenvalues equation
\begin{equation}
-\left[\frac{1}{\sin\theta}\frac{\partial}{\partial\theta}
\left(\sin\theta\frac{\partial}{\partial\theta}\right)+\frac{1}{{\sin}^2\theta}
\frac{{\partial}^2}{\partial{\phi}^2}\right]Y_{\ell m}=\ell(\ell+1)Y_{\ell m},
\label{42}
\end{equation}
while for the radial part, after defining $u_E(r)=rR(r)$, we have
\begin{equation} 
-\frac{{\hbar}^2}{2m}u_{E,r,r}+\left[V(r)+\frac{\ell(\ell+1){\hbar}^2}{2m r^2}\right]u_E=E u_E,\;\;u_E(r=0)=0.
\label{43}
\end{equation}
Consider now the free case by setting $V(r)=0$. The (\ref{43}) becomes
\begin{equation} 
-{\hbar}^2 u_{E,r,r}+\frac{\ell(\ell+1){\hbar}^2}{r^2} u_E=P^2 u_E.
\label{kk}
\end{equation}
In the (\ref{43}) the Planck constant arises thanks to the usual form of the impulse
operator $\mathbf{P}$ as $\mathbf{P}=-\imath\hbar\nabla$. To make the analogy between the (\ref{43}) and the (\ref{f3}) 
(and (\ref{23})) we can write the (\ref{43}) as
\begin{equation}
-u_{E,r,r}+\frac{\ell(\ell+1)}{r^2}u_E=k^2u_E,\;\;\;k=\sqrt{\frac{2mE}{{\hbar}^2}}.
\label{44}
\end{equation}
Since for the (\ref{44}) $E=\frac{P^2}{2m}$, we have $k=\frac{|P|}{\hbar}$. 
We can now rewrite equation (\ref{f3}) as 
\begin{equation}
-Z_{\ell m,r,r}^{(p)}+\frac{\ell(\ell+1)}{r^2}Z_{\ell m}^{(p)}=\frac{{\omega}^2}{c^2}Z_{\ell m}^{(p)}.
\label{45}
\end{equation}
Thanks to the (\ref{45}) and (\ref{44}) we can write $|P|=\frac{\hbar\omega}{c}$. This means that the equation governing the polar
\footnote{For the axial perturbation the analogy is only with the radial eqution (\ref{23}).} 
perturbations can be interpreted as a wave function for free massless gravitons with energy $E=\hbar\omega$ and momentum
$|P|=\frac{E}{c}$.
As a consequence, the way we have to take evident the quantum nature of gravitons is to confine gravitons in a finite
(spherical in our context) box. As preliminarily discussed in the introduction, this is not a simple task. Various attempts can be found in the 
literature (see for example \cite{15,16,17,18,19,20,21}) with opposite answers. In this paper we use the pragmatic view present in
\cite{15} avoiding the issues of a practical realization. Quite
reasonably \cite{21}, we could think to a gravitational radiation in thermal equilibrium with the (hard) wall of a spherical box at a given temperature $T$. Hence, we have a potential $V(r)$ that is zero inside the box and practically infinity at the boundary
$R$.

To start with, we must specify the solution of equation (\ref{45}). As well known \cite{14}, the regular solution for 
$L_{\ell m}(r,\omega)$ can be obtained in terms of regular Bessel functions $j_{\ell}(kr)$ with $k=\frac{\omega}{c}$.
Thus we have $L_{\ell m}(r,\omega)=A_{\ell}\;j_{\ell}(kr),A_{\ell}\in\mathbb{R}$. The regular Bessel functions for $r\rightarrow 0$ look like
$j_{\ell}(kr)\sim {{(kr)}^{\ell}}/(2\ell+1)!!$, while for large values of $kr$ we have
\begin{equation}
j_{\ell}(kr)\simeq\frac{1}{kr}\cos\left[kr-\frac{(\ell+1)\pi}{2}\right].
\label{51}
\end{equation}
It should be noticed that the (\ref{51}) is a good approximation also for $kr > \ell$.
To confine our gravitational radiation within a spherical box of radial radius $R$, we must impose the boundary condition
$L_{\ell m}(R,\omega)=0$: this condition is a Dirichlet boundary condition and in fact the confining box
is a Dirichlet boundary condition for the metric function $Z_{\ell m}$.

As a consequence, with the help of (\ref{51}), the following discrete spectrum for the so trapped gravitons 
does arise:
\begin{equation}
{\omega}_{\ell n}\simeq\frac{c}{2R}\left(2+\ell+2n\right)\pi,\,\;\ell\geq 2,\;\;n\in\mathbb{N}.
\label{52}
\end{equation}
As expected, finite size effects lead to the manifestation of a quantum behavior for gravitons. The classical behavior is restored
for $R\rightarrow\infty$ where a strictly continuum spectrum arises. 
Note that the allowed frequencies look like the ones of an harmonic oscillator but with the further
dependence on the Legendre index $\ell$. 
As the usual harmonic oscillator, the ground state of (\ref{52}) has a non-vanishing energy $E_{0\ell}$ given by
$E_{\ell 0}=\hbar{\omega}_{\ell 0}=\hbar\frac{c(\ell+2)\pi}{2R}$.
In the following, we will discuss how to interpret equation (\ref{52}).
Thanks to the similarity depicted above, we has been able to obtain a physically reasonable expression for the spectrum of trapped gravitons.
By forcing this analogy, 
we are tempted to promote the indices $\{\ell,m,n\}$ to quantum numbers. However, note that
formula (\ref{52}) is independent from the azimuthal index $m$ and as a result the energy levels are degenerate with a degeneration factor
given by $2\ell+1$. However, it should be noted that the Regge-Wheeler and Zerilli equations (\ref{f3}) and (\ref{23}) have been obtained 
by perturbing a spherically symmetric vacuum region. As a consequence, for any value of the Legendre index $\ell$ we can have 
quadrupolar ($\ell=2$), sextupolar ($\ell=3$) perturbations. To any kind of perturbations we can associate a different species of gravitons, i.e.
we can have a given harmonic oscillator spectrum for any given fixed $\ell$: thus we can have
quadrupolar, sextupolar... gravitons. Hence, only the index $n$ could be promoted to a quantum index, whereas the index $\ell$ can be seen as
a species index. This interpretation of the index $\ell$ could be of interest, for example, in relation to the so named 'species problem'
\cite{s1}. In fact, 'ordinary' entropy depends on the number of species, while the black hole entropy is species-independent.
In our case, we have infinity 'species' of gravitons 
\footnote{Only massless particles can stably survive inside a black hole.}
that must be summed up in order to obtain the geometric expression of the black hole entropy.
In the following, we adopt this point of view. Nevertheless, for completeness of presentation, in section 7 we study the 
interpretation with all
the indices $\{\ell,m,n\}$ promoted to quantum numbers. 

\section{Statistical mechanics of gravitons in a box}

Our starting point \cite{hua} is the canonical partition function $Z_{\ell}$ for the discrete spectrum (\ref{52}) with the summation over the quantum index $n$:
\begin{equation}
Z_{\ell}=\sum_{n=0}^{\infty} e^{-\beta\hbar{\omega}_{\ell n}},\;\;\;\beta=\frac{1}{K_B T}.
\label{55}
\end{equation}
First of all, we study the case  of the partition function (\ref{55}) with a given Legendre index $\ell$ .
With a trivial algebra we obtain
\begin{equation}
Z_{\ell}=\frac{e^{-\left[\frac{c\beta}{2R}(\ell+2)\pi\hbar\right]}}
{1-e^{-\left[\frac{c\beta\pi\hbar}{R}\right]}}.
\label{56}
\end{equation}
The Helmholtz free energy $F_{\ell}$ is $F_{\ell}=-K_B T\ln(Z_{\ell})$. 
Hence, the numerator in the (\ref{56}) is a constant independent from
$\beta$ and for the entropy $S_{\ell}$ we obtain the one of a discrete harmonic oscillator with 
angular frequency $\omega=c\pi/R$. For the mean energy $<E_{\ell}>$ we obtain
\begin{equation}
<E_{\ell}>=\frac{c}{2R}(\ell+2)\pi\hbar+
\frac{c\pi\hbar\;e^{-\left[\frac{c\beta\pi\hbar}{R}\right]}}{R\left({1-e^{-\left[\frac{c\beta\pi\hbar}{R}\right]}}\right)}.
\label{57}
\end{equation}
By denoting	with ${\omega}_g=c\pi/R$, expression (\ref{57}) becomes
\begin{equation}
<E_{\ell}>=\left(<n>+\frac{(\ell+2)}{2}\right)\hbar{\omega}_g.
\label{58}
\end{equation}
It is intriguing to note that, after formally setting $\ell=1$ in (\ref{58}), we obtain the usual formula for an ordinary three dimensional 
harmonic oscillator\footnote{This is what we expect, thanks to its dipolar nature, for the electromagnetic radiation.}.
The term $\frac{(\ell+2)\hbar{\omega}_g}{2}$ in (\ref{58}) is nothing else but the non-vanishing energy of the ground state.
In the Planck-like derivation, the ground state energy is 
usually set to zero: in our case this implies that the term dependent on $\ell$ is absent. In fact, this term 
\footnote{This represents the zero point energy.}
is typically subtracted to $<E>$. In this way, the usual Planck distribution emerges with the mean number occupation $<n_{\omega}>$ for the level
with energy $\hbar\omega$ given by
\begin{equation}
<n_{{\omega}_g}>=\frac{1}{e^{\beta\hbar{\omega}_g}-1},
\label{59}
\end{equation}
that is the Planck distribution. It may be argued that if we consider a general gravitational perturbation, then a summation over 
$\ell$ and $m$
must be performed. As stated at the end od section 3, in the following we consider the index $\ell$ as representing the species number and we leave
the case of the summation over $m$ at the appendix.
We must thus calculate the full partition function after marginalization of $\ell$. We obtain:
\begin{equation}
Z_g=\sum_{\ell=2}^{\infty}\sum_{n=0}^{\infty} e^{-\beta\hbar{\omega}_{\ell n}}=
\frac{e^{-\left(\frac{2c\pi\beta\hbar}{R}\right)}}
	{\left[1-e^{-\left(\frac{c\pi\beta\hbar}{2R}\right)}\right]}
	\frac{1}{\left[1-e^{-\left(\frac{c\pi\beta\hbar}{R}\right)}\right]}.
\label{60}
\end{equation}
As a consequence of the (\ref{60}), after calculating the Helmholtz energy $F_g=-K_B T\ln(Z_g)$, we see that the entropy is the one of two
harmonic oscillators with angular frequencies ${\omega}_1=c\pi/(2R)$ and ${\omega}_2=2{\omega}_1$. For the mean energy 
$<E_g>$ we obtain in the space of frequencies ${\nu}$
\begin{equation}
<E_g>=h{\nu}_0+\frac{h{\nu}_1}{\left[e^{\beta h{\nu}_1}-1\right]}+\frac{h{\nu}_2}{\left[e^{\beta h{\nu}_2}-1\right]},\;
{\nu}_0=\frac{c}{R},\;{\nu}_1=\frac{c}{4R},\;{\nu}_2=\frac{c}{2R}.
\label{61}
\end{equation} 
Note that the ground state energy $h{\nu}_0$ can be obtained from the one given in (\ref{57}), i.e.
$\frac{c}{2R}(\ell+2)\pi\hbar$, by setting $\ell=2$: this could be a manifestation of the quadrupolar nature of the gravitational radiation.
After subtracting this term we obtain
\begin{equation}
<E_g>=\frac{h{\nu}_1}{\left[e^{\beta h{\nu}_1}-1\right]}+\frac{h{\nu}_2}{\left[e^{\beta h{\nu}_2}-1\right]}.
\label{62}
\end{equation} 
The (\ref{62}) is a summation of two discrete Planck distributions with proportional frequencies, i.e. 
${\nu}_2=2{\nu}_1$. In this regard, gravitons seem to differentiate from Planck distributions of photons in thermal 
equilibrium within a cavity. The term with ${\nu}_1$ is a consequence of the summation over the Legendre index $\ell$ and thus 
can be seen as the mean contribution to $E_g$ related to the superposition of all possible kind of gravitational perturbations
(quadrupolar, sextupolar...). More generally, 
a gravitational radiation in thermal equilibrium within a cavity can be seen as the superposition of $N$ 
harmonic oscillators with fundamental
frequencies ${\nu}_1$ and ${\nu}_2=2{\nu}_1$ with partition function $Z_T$ \cite{hua}
\begin{equation}
Z_T=Z_g^N.
\label{63}
\end{equation}
All these facts imply a difference with the ordinary electromagnetic radiation in a cavity where the term with 
${\nu}_1$ is absent. To this purpose, it should be always held fixed in mind the fact that electromagnetic radiation is typically dipolar, while
the gravitational one is quadrupolar.  

\section{Thermodynamics of gravitons}

First of all, thanks to the (\ref{63}), for a system of $N$ oscillators we have:
\begin{equation}
Z_T=
{\left(\frac{e^{-\left(\frac{2c\pi\beta\hbar}{R}\right)}}
{\left[1-e^{-\left(\frac{c\pi\beta\hbar}{2R}\right)}\right]}
\frac{1}{\left[1-e^{-\left(\frac{c\pi\beta\hbar}{R}\right)}\right]}\right)}^N.
\label{64}
\end{equation}
The internal energy $<E_T>=U$ can be derived from
(\ref{64}) in the usual way as $-{(\ln Z_T)},{\beta}$:
\begin{equation}
U=\frac{c\pi\hbar N}{2R\left[e^{\frac{\beta c\pi\hbar}{2R}}-1\right]}+\frac{c\pi\hbar N}{R\left[e^{\frac{\beta c\pi\hbar}{R}}-1\right]}.
\label{65}
\end{equation}
Note that in the limit for $R\rightarrow\infty$ (at fixed $T$ or more precisely for $\hbar\rightarrow 0$) we have $U\rightarrow 2N K_B T$.
By inspection of the (\ref{65}) we see that, in our discretized system, the internal energy $U$ is a function of the thermodynamic
temperature $T$ and the radius $R$. Hence, we must pay care to use Maxwell's relations involving partial derivatives with respect to the volume 
$V=4/3\pi R^3$. As an example, by the first law we have $T dS=dU+PdV$. Since $U=U(R,T)$, it is not more true that, for example
$\frac{\partial S}{\partial V}=\frac{P}{T}$, with $P$ the pressure.\\   
To start with, we thus calculate the entropy by using the usual definition
\begin{equation}
S=-K_B\sum_{\ell=2}^{\infty}\sum_{n=0}^{\infty}P_{n\ell}\ln(P_{\ell,n}),\;\;
P_{n\ell}=\frac{e^{-\beta E_{\ell n}}}{Z_T}.
\label{66}
\end{equation}
After a simple algebra we obtain
\begin{equation}
TS=NK_B T\ln(Z_g)+U,
\label{67}
\end{equation}
and with $F=U-TS$ we regain the usual relation $F_T=-NK_B T\ln(Z_g)$. As a consequence, since $dF_T=-SdT-PdV$ certainly we have
that $\frac{\partial F_T}{\partial T}=-S$ and $\frac{\partial F_T}{\partial V}=-P$. 
Hence, for $S$  we obtain
\begin{eqnarray}
& & S=-N K_B\left[\ln\left(1-e^{-\frac{X}{2}}\right)+\ln\left(1-e^{-X}\right)\right]+\label{68}\\
& & +\frac{c\pi\hbar N e^{-\frac{X}{2}}\left[1+3 e^{-\frac{X}{2}}\right]}{2TR\left(1-e^{-X}\right)},\;\;
X=\frac{c\pi\beta\hbar}{R},\nonumber
\end{eqnarray}
and for the pressure $P$
\begin{equation}
P=\frac{c\pi N\hbar\left[e^{-\frac{X}{2}}+3e^{-X}\right]}{2R^2\left(1-e^{-X}\right)}\frac{dR}{dV},
\label{69}
\end{equation}
with $dR/dV=3^{-1}{(3/(4\pi))}^{1/3} V^{-2/3}$. It is a simple matter to verify that, thanks to the formula (\ref{65}), we regain the usual relation for a
radiation field $PV=U/3$. In the limit for $\hbar\rightarrow 0$ (classical limit) we obtain in particular
$PV=2NK_B T/3=U/3$.\\ 
Summarizing, in this section we have shown that the application of the (\ref{52}) for trapped gravitons
in the semiclassical limit, by considering the Legendre index as 
labeling the species of gravitons, leads to well defined thermodynamic quantities together with the expected equation of state 
(\ref{69}) suitable for a radiation field. In the appendix we performed the same calculations of this section but with the degeneracy factor
$(2\ell+1)$ taken into account.\\ 
In the next section we apply the results of this section in order to obtain the Schwarzschild
black hole entropy from a statistical point of view.

\section{An application: the Schwarzschild black hole entropy}
 
An important issue of modern physics is the statistical understanding of the black hole entropy $S_{BH}=K_B A_h/L_P^2$, i.e. the calculation 
of the internal degrees of freedom of a black hole. A lot of attempts are present in the literature (see for example
\cite{22,23,24,25,26,27,27b,27bb} and references therein). Typically,
in many approaches black hole entropy arises from a proposed quantum gravity theory
(string theory, loop quantum gravity) and then the validity of these derivations is related to the unproven used theory.
Other approaches use excitation modes propagating near the horizon \cite{22,23,24}. In particular, in \cite{22} and 
\cite{24} the authors suppose that the black hole entropy is the entanglement between internal and external (with respect to the horizon)
degrees of freedom, but the expression so obtained is ultraviolet divergent.
Generally, the main obstacle to get a statistical derivation of the Bekenstein-Hawking entropy is due to the presence of the Planck length
$L_P$, i.e. how to obtain the constant $L_P$ without using a prosed quantum gravity theory. In fact, in many attempts the proportionality
$S\sim A$ is obtained but the constant of proportionality is introduced by hand.\\
In the following, we show that, thanks to the formula (\ref{52}) and to the fact that we do not perform the continuum (classical) limit
by retaining $\hbar\neq 0$, the Bekenstein-Hawking entropy formula can be obtained in a clear and simple way. In practice, since the formula $S_{BH}=K_B A_h/L_P^2$ is semi-classical, we use
a discrete version of the canonical ensemble distribution.\\
We note that black hole is a natural arena to apply the results of sections 4 and 5 since nothing can escape from a black hole and 
the event horizon can be considered as a perfect hard wall. Inspired by perturbative quantum field theory \cite{E1},
we obtained the frequency formula
(\ref{52}) by perturbating a spherical region in a Minkowski spacetime, 
while the Schwarzschild geometry is not flat.
Nevertheless, many arguments can be invoked to justify the application of this section.

In the situation considered in the sections above gravitons are trapped in a spherical region and the (\ref{52}) it gives the spectrum of these
gravitons. We may suppose the interior of a black hole as composed of a gas of massless excitations provided by gravitons. These gravitons
with their frequencies (energy) spectrum (\ref{52}) generate the curvature inside the event horizon at $r=R_s$.
The condition that these so obtained gravitons generate a black hole is provided by equating the internal energy of this gas with the ADM
mass-energy $Mc^2$ of the black hole\footnote{This can be obtained by putting a sufficiently large amount $N$ of 
oscillators within a given radius $R$, as shown by equation (\ref{71})}. This is a standard 'empirical' procedure that avoids
the shortcomings due to the lack of a complete quantum gravity theory.\\  
Moreover, there exists evidence that
(see for example \cite{22,23,24}) excitation modes inside the Schwarzschild black hole propagates near the event horizon.
This fact maybe a consequence of the holographic principle  \cite{28,29}
since the internal degrees of freedom are in some sense 'projected'
onto the black hole event horizon, i.e. the holographic screen. It is a known fact that the metric near the horizon is 'almost'
flat (see for example \cite{a2}). Near the horizon of a Schwarzschild black hole, the metric looks like the one of
a flat Minkowski spacetime times a sphere of radius $R_s$ given by the one of the event horizon, i.e. 
$R_s=2GM/c^2$ (Rindler spacetime). 
An experimenter very close to the event horizon cannot distinguish between a Rindler coordinate system
in a flat spacetime with the black hole Schwarzschild spacetime 
\footnote{The Hawking temperature is in fact defined by the surface gravity on the event horizon.}
and he is legitimate to adopt usual thermodynamics.
This point of view also inspired the works based on the concept of entanglement entropy
(see again the nice paper \cite{E1}).\\
Moreover, In the study of the thermodynamics of (at least) static black holes, for its volume $V_h$ \cite{a7,33} it is appropriate
(named in this case thermodynamic volume) the usual Minkowskian form $V_h=4\pi R_s^3/3$, irrespective of the fact that inside the 
horizon the spatial coordinate $r$ becomes timelike and the time coordinate $t$ spacelike and the usual definition of a volume is rather problematic. 
Summarizing, also the holographic principle conspires to elevate the formula (\ref{52}) as a physically reasonable expression  
for the spectrum of trapped gravitons inside the event horizon with the black hole degrees of freedom stored
near the horizon. In the following, we show that this is encoded by the behavior of $N$ that looks like $A\sim R_s^2$ rather than the 
volume $V$ of the event horizon.

To start with, we must specify the internal energy $U$. For a Schwarzschild black hole, as stated above,
this is done by its rest ADM energy, i.e.
$U=Mc^2$, where $M$ is the ADM black hole mass. By the usual relation $2GM/c^2=R$ we obtain $U=c^4 R/(2G)$. The first step is the determination
of the behavior of $N$ and $T_i$. This can be done by considering a suitable thermodynamic limit. 
At a first look, we can use the standard set up $N\sim R^3$. With this choice
and at a classical level with with the temperature independent from the Planck constant, 
formula (\ref{65}) implies that\footnote{This can be understood by the fact that 
in the limit $\hbar=0$ we have $U=2NK_BT$.} $T_i\sim 1/R^2$. When the expression $T_i\sim 1/R^2$ is inserted in the entropy formula
(\ref{68}), in the limit $R\rightarrow\infty$ we obtain $S\rightarrow 0$. This result is obviously 
not acceptable. Moreover, the constraint $U\sim R,\;\forall R>0$ rule out this possibility.
The only non-classical possibility leading to $S\sim A$
is that, according to the holographic principle \cite{28,29}, we have 
$N\sim R^2$. Since for a black hole $U\sim R$, thanks to the (\ref{65}), for the internal temperature $T_i$ of a black hole we must have
\begin{equation}
T_i=\alpha T_h=\frac{\alpha c\hbar}{4\pi K_B R},\;\;\alpha\in (0,\infty).
\label{70}
\end{equation}
Formula (\ref{70}) means that the internal temperature is proportional to the one of the event horizon. There is not a priori reason to assume 
$\alpha=1$. In fact, the temperature $T_h$ is proportional to the surface gravity parameter on the event horizon and is normalized in such a way that the norm of the timelike killing vector is the unity at spatial infinity. However, for an unlucky observer inside the black hole the outside
spatial infinity is not attainable and the usual normalization is not available. Hence, the result that $\alpha\neq 1$ is 
physically reasonable, although we expect $\alpha$ of the order of unity.\\
As a consequence, thanks to the (\ref{70}) and using $U=c^4 R/(2G)$, the (\ref{65}) it gives
\begin{equation}
R=\sqrt{\pi N}L_P\sqrt{\frac{1}{\left[e^{\frac{{2\pi}^2}{\alpha}}-1\right]}+
	\frac{2}{\left[e^{\frac{4{\pi}^2}{\alpha}}-1\right]}}.
\label{71}
\end{equation}
Note the presence of $L_P$ in the (\ref{71}) that is a consequence of the presence of $\hbar$ in (\ref{65}).
Finally, with 
(\ref{71}) the entropy (\ref{68}) becomes
\begin{eqnarray}
& & S=K_B Y(\alpha)\frac{A_h}{4L_P^2},\;\;A_h=4\pi R^2,\label{72}\\
& & Y(\alpha)=\frac{b}{\alpha\pi^2\left(3+e^{\frac{2\pi^2}{\alpha}}\right)}\nonumber\\
& & b=-\alpha e^{\frac{4\pi^2}{\alpha}}\ln\left(1-e^{-\frac{2\pi^2}{\alpha}}\right)-
\alpha e^{\frac{4\pi^2}{\alpha}}\ln\left(1-e^{-\frac{4\pi^2}{\alpha}}\right)+6\pi^2+\nonumber\\
& & +2\pi^2 e^{\frac{2\pi^2}{\alpha}}+\alpha \ln\left(1-e^{-\frac{2\pi^2}{\alpha}}\right)+
 \alpha\ln\left(1-e^{-\frac{4\pi^2}{\alpha}}\right).  
\end{eqnarray} 
It is a simple matter to see that the equation $Y(\alpha)=1$ has (remarkable) 
a unique solution approximatively given by
$\alpha\simeq 2.225$. First of all, it should be noted that formula (\ref{52}) is an approximation, but a good approximation for the roots of
$j_{\ell}$. In particular, the maximum error is for the first zero that is of the order of the $3\%$. This error quickly decreases for the
remaining zeros of $j_{\ell}$ and as a result  
a small correction to the value obtained for $\alpha$ arises with its 'true' value of the order of
$2$. Physically, this factor can be interpreted as due to the fact that the temperature $T_h$ is measured by an observer at spatial infinity, while
for an observer within the event horizon the outside spatial infinity is not available. Moreover, note that, thanks to
the (\ref{69}), a non-vanishing pressure is present inside the horizon due to the gravitons radiation pressure. As a consequence, since
$T_idS_{BH}=dU+PdV_h$ and $S$ and $U$ are left unchanged\footnote{Remember that for the first law
	at $R=2GM/c^2$ we have $T_hdS_{BH}=dU$}, we must have $T_i\neq T_h$. This pressure $P$ looks like $P\sim 1/R^2$, that is a further 
indication that gravitons behave holographic inside the black hole. This is an interesting result that solves the problem concerning the presence 
of a work term in the usual black hole first law of thermodynamics. An observer measures the Hawking flux at spatial infinity with a temperature
$T_h$ and thus is forced to associate to the black hole an entropy $S_{BH}\sim A_h/4$, a temperature $T_h$ 
(the temperature he/her perceives) and an internal energy given by the ADM mass. 
For an internal observer the situation is different. He perceives the temperature of the gravitons gas present within the horizon
and also the pressure generated by this gas. Both agree that the entropy is $S_{BH}\sim A_h/4$, but the 'internal' point of view in a more usual way since he can see gravitons inside the black hole. Hence the usual form of the first law $T_h dS_{BH}=c^2 dM$ is a consequence of the
lack of knowledge of the nature of the degrees of freedom inside the event horizon of the external observer.

It should be noticed that our result has been obtained only by requiring that the internal energy is given by the total ADM mass-energy of the black hole. This is a physically sound assumption since the internal energy is the energy necessary to 'create' a given system and
the ADM mass is the total energy of the black hole.

As a final consideration for this section, we study the ultraviolet term $R\rightarrow 0$. In the expression (\ref{65}) we have subtracted, as 
usual, the 'ultraviolet' term $U_0=2\pi c\hbar N/R$. 
This term is obviously diverging for $R\rightarrow 0$ at fixed finite $N$ and it is thus expected 
to play a role when the Planck length $L_P$ is reached. This term becomes dominant when $R\sim L_P$. 
By inserting this term in the (\ref{65}), we see that the leading term of the (\ref{71}) for 
$R\sim L_P$ is given by:
\begin{equation}
R\simeq 2\sqrt{\pi N}\;L_P.
\label{q1}
\end{equation}
At the Planck length it is expected that 
the spacetime geometry is not more given by the usual one but substituted with a non-commutative 
spacetime thus evidencing a quantum structure for the 
spacetime, i.e. a quantum spacetime \cite{a3,a4}.
For a quantum spacetime, uncertainty relations involving coordinates come into action (see \cite{a3} and the generalization
in \cite{a5,a6}). In particular, for spherical geometries, a minimal uncertainty $\Delta R$
for the radial coordinates does arise of the order of the 
Planck length \cite{a5}, i.e. $\Delta R\geq \sqrt{3} L_P$: the expression (\ref{q1}) is in agreement with this inequality. In particular,
we may think that the minimum radius $R_{min}$ can be obtained in (\ref{q1}) by taking a single ($N=1$) oscillator thus obtaining
$R_{min}\simeq 2\sqrt{\pi}\;L_P$. This suggests that a quantum spacetime in a state with spherical shape can be seen as composed of 
spherical balls each filled with a single 
harmonic oscillator made of a single oscillating graviton.

\section{A summation over $m$}

In section 3 the analogy between the Regge-Wheeler and Zerilli equations
with the Schrodinger one allowed us to physically motivate the 
(\ref{52}) as representing the spectrum for trapped gravitons. By forcing this interpretation, we can promote the indices
$\{\ell,m,n\}$ to quantum numbers representing quantum states.
In this section we study the case where the summation over the azimuthal index $m$ is performed and as a consequence instead of the 
(\ref{60}) we have 
\begin{equation}
Z_g=\sum_{\ell=2}^{\infty}\sum_{m=-\ell}^{m=+\ell}\sum_{n=0}^{\infty} e^{-\beta\hbar{\omega}_{\ell n}}=
\sum_{\ell=2}^{\infty}\sum_{n=0}^{\infty}(2\ell+1)e^{-\beta\hbar{\omega}_{\ell n}}.
\label{a1}
\end{equation}
To perform the summation (\ref{a1}), we can use the following usual formula that is fulfilled for a geometric series for 
$x>0$;
\begin{equation}
f(x)=\frac{1}{1-e^{-x}}=\sum_{i=0}^{\infty}e^{-ix},\;\frac{df(x)}{dx}={\left(\frac{1}{1-e^{-x}}\right)}_{,x}=
-\sum_{i=1}^{\infty}ie^{-ix}.
\label{a2}
\end{equation}
With the (\ref{a2}) we obtain for $Z_g$, with $X=\frac{c\pi\beta\hbar}{R}$,
\begin{equation}
Z_g=\frac{e^{-2X}(3-2 e^{-\frac{X}{2}})}{(1-e^{-X}){(1-e^{-\frac{X}{2}})}^2}.
\label{a3}
\end{equation}
Formula (\ref{a3}) contains extra terms with respect to (\ref{60}). For the total free energy $F_T$
we obtain:
\begin{equation}
F_T=-K_BN T\left[-2X+\ln\left(3-2e^{-\frac{X}{2}}\right)-\ln\left(1-e^{-X}\right)-
2\ln\left(1-e^{-\frac{X}{2}}\right)\right].
\label{a4}
\end{equation}
For the entropy $S$ we obtain:
\begin{equation}
S=-\frac{F_T}{T}+\frac{c\pi N\hbar e^{-\frac{X}{2}}}{TR}
\frac{\left[2+4 e^{-\frac{X}{2}}-3e^{-X}\right]}{\left(3-2 e^{-\frac{X}{2}}\right)\left(1-e^{-X}\right)}.
\label{a5}
\end{equation}
First of all, we can study the internal energy $U$ (after subtracting the zero energy value):
\begin{equation}
U=\frac{c\pi N\hbar}{R}\left[-\frac{1}{\left(3e^{\frac{X}{2}}-2\right)}+\frac{1}{\left(e^X-1\right)}
+\frac{1}{\left(e^{\frac{X}{2}}-1\right)}\right].
\label{a6}
\end{equation}
Note the negative term in the (\ref{a6}). Nevertheless, we have $U>0\;\forall X>0$.
In the limit $X\rightarrow 0$ ($T\rightarrow\infty$ at fixed $R$ or $R\rightarrow\infty$ at $T$ fixed) we have
$U\rightarrow 3NK_B T$ ($2NK_BT$ for the (\ref{65})), whereas for
$X\rightarrow\infty$ we always have $U\rightarrow 0$. 

Also for the (\ref{a1}) we can apply the same arguments of section 6. In particular, the (\ref{70}) is still valid and the
(\ref{71}) becomes:
\begin{equation}
R=\sqrt{\pi N}L_P\sqrt{\frac{2}{\left[e^{\frac{{2\pi}^2}{\alpha}}-1\right]}+
	\frac{2}{\left[e^{\frac{4{\pi}^2}{\alpha}}-1\right]}-\frac{2}{\left[3e^{\frac{{2\pi}^2}{\alpha}}-2\right]}}.
\label{a7}
\end{equation}
By following the same arguments of section 6, we found that also with the (\ref{a1}) we obtain the 
Bekenstein-Hawking entropy only for a value, i.e. $\alpha\simeq 8.48$.

Summarizing, the summation over $m$ leads to corrections with respect to the formulas of sections 4-6, but the physical features remain substantially unchanged.

\section{Conclusions and final remarks}

In this paper, our attention focused on the analogy between the Regge-Wheeler 
(\ref{23}) and Zerilli (\ref{f3}) equations with the Schrodinger one. Obviously, the key question is under what conditions
$Z^{(a,p)}$ behaves, in practice, like a quantum wavefunction. Since we described free massless gravitons, equations
(\ref{23}) and (\ref{f3}) are always 'classical'.
A reasonable way to see the quantum behavior of gravitons is to  
confine the gravitational radiation in a finite spherical box of radial radius $R$ and thus impose the appropriate 
(Dirichlet) boundary conditions,
i.e. $Z_{\ell m}^{(a,p)}(t,R)=0$. As a result of this condition, a discrete spectrum arises dependent on the Legendre index
$\ell$ and the integer $n$ of usual harmonic oscillators. The physical situation can be assimilated to the one of a certain gravitational
radiation \cite{21} in thermal equilibrium within a given cavity. The index $\ell$ can be as well be interpreted as representing
the species of gravitons present after perturbation. In this case, 
by calculating the canonical partition function, as a result of a
summation over $\ell$ and $n,$ the so obtained distribution is a superposition of two Planck distributions 
(\ref{62}) with proportional frequencies. This is an interesting result since differs from the single Planck distribution of photons in thermal equilibrium within
a cavity.\\ 
The other possibility is to promote all the indices (not only $n$) $\{\ell,m,n\}$ to quantum numbers and a summation
with respect to $m$ becomes necessary. This case is studied in section 7 with formulas similar to the ones obtained in sections
3-6 and with physical features substantially unchanged.

As an interesting application, we used our results to obtain the Bekenstein-Hawking entropy in a simple and physically clear way. 
The only assumption has been that the internal energy $U$ of the black hole is provided by its ADM rest energy, a very reasonable assumption.
As a result, in order to have consistency with the behavior $U\sim R$ and an entropy proportional to the area of the event horizon,
we must have $N\sim R^2$ and as a consequence we obtain 
for the internal temperature $T_i$ of the black hole $T_i\simeq 2.225 T_h$ ($T_i\simeq 8.48 T_h$ by summing up with respect to $m$). 
These rather intriguing results have a simple physical
interpretation. The internal degrees of freedom of a black hole are provided by gravitons that can be seen as an ensemble of $N$
harmonic oscillators. Every single graviton contributes to the total entropy by a quantity proportional to the Planck area
$A_P=4\pi L_P^2$, in agreement with the holographic nature of the black hole entropy. 

Concerning the approach used in this paper, some analogy can be done with the papers quoted in 
\cite{E3,CC1,CC2,CC3}. There, the author uses the correspondence between Hawking radiation and black hole quasi-normal modes
that are in turn interpreted as quantum levels in a semi-classical model inspired by the Bohr approach used in 1913.
The interesting consequences of these setups are that black hole entropy is proportional to the black hole excited states and
quite remarkably represent a possible reasonable solution to the black hole information paradox. 
Since we attempt to describe the interior of the black hole near the horizon, while in the papers \cite{E3,CC1,CC2,CC3}
the exterior near horizon region is considered, our approach could be seen as complementary to the one present in
the quoted above papers.\\
Moreover, note that the discrete spectrum formula (\ref{52}) for a graviton of species $\ell$
can be rewritten in the form
\begin{equation}
{\omega}_{\ell n}\frac{R}{c}\simeq \left(1+\frac{\ell}{2}+n\right)\pi
\label{af}
\end{equation}
that, apart from the term $\left(1+\frac{\ell}{2}\right)\pi$, is related to the famous
Bohr quantization rule $\oint pdq=nh$ for the coordinate $q$ and its conjugate variable $p$ in the form 
\begin{equation}
\oint pdq=\left(1+\frac{\ell}{2}+n\right)\pi h,
\label{af2}
\end{equation}
with $q=R$ and $p=\hbar{\omega}_{\ell n}/c$.

As a final consideration, notice that to obtain our results we have not invoked particular quantum gravity theories, 
and for this reason this makes our result in some sense 'phenomenological' and thus physically sound. A possible further 
development of the argument presented in this paper is to explain in a statistical sense the recently proposed generalization for the black
hole entropy in a cosmological context \cite{30,31,32,33} and in particular the recent proposal 
\cite{34} concerning the nature of the cosmological constant in terms of 'slow' gravitons.

\section*{Acknowledgements}  
I would like to thank Luca Tomassini for useful discussions concerning non commutative spacetimes and the anonymous referee who
suggested the references \cite{E3,CC1,CC2,CC3} in relation with the approach used in this paper.

\end{document}